\documentstyle[epsfig]{aipproc}

\begin{document}
\title{First results from \\ the BOOMERanG experiment}

\author{P. de Bernardis$^{1}$,
P.A.R.Ade$^{2}$,  
J.J.Bock$^{3}$, 
J.R.Bond$^{4}$, 
J.Borrill$^{5,6}$, 
A.Boscaleri$^{7}$, 
K.Coble$^{8}$, 
B.P.Crill$^{9}$,
G. De Gasperis $^{10}$, 
G. De Troia$^{1}$, 
P.C.Farese$^{8}$, 
P.G.Ferreira$^{11}$, 
K.Ganga$^{9,11}$, 
M.Giacometti$^{1}$, 
E.Hivon$^{9}$, 
V.V.Hristov$^{9}$, 
A.Iacoangeli$^{1}$, 
A.H.Jaffe$^{6}$, 
A.E.Lange$^{9}$, 
L.Martinis$^{13}$, 
S.Masi$^{1}$, 
P.Mason$^{9}$, 
P.D.Mauskopf$^{14}$, 
A.Melchiorri$^{1}$, 
L.Miglio$^{1,15}$, 
T.Montroy$^{8}$, 
C.B.Netterfield$^{15}$, 
E.Pascale$^{7}$, 
F.Piacentini$^{1}$, 
D.Pogosyan$^{4}$, 
F.Pongetti$^{16}$,
S.Prunet$^{4}$, 
S.Rao$^{16}$, 
G.Romeo$^{16}$, 
J.E.Ruhl$^{8}$, 
F.Scaramuzzi$^{13}$, 
D.Sforna$^{1}$, 
N.Vittorio$^{10}$ 
 }
\address{$^1$ Dipartimento di Fisica, Universit\'a di Roma La Sapienza,
Roma, Italy;
$^2$ Dept. of Physics, Queen Mary and Westfield College, London, UK; $^3$
Jet Propulsion Laboratory, Pasadena, CA, USA; $^4$ CITA University of
Toronto, Canada; 
$^5$ NERSC-LBNL, Berkeley, CA, USA; $^6$ Center for Particle Astrophysics, 
Univ. of California at Berkeley, USA; $^7$ IROE - CNR, Via Panciatichi 64,
50127 Firenze, Italy; $^8$  Department of Physics, Univ. of California at
Santa Barbara, USA; 
$^9$ California Institute of Technology, Pasadena, USA; $^{10}$
Dipartimento di Fisica, Universit\'a di Roma Tor Vergata, Roma,
Italy; $^{11}$ Astrophysics, University 
of Oxford, UK; $^{12}$ PCC, College de France, Paris, France; $^{13}$ ENEA
Centro 
Ricerche di Frascati, Italy ; $^{14}$  Physics and Astronomy Dept, Cardiff
University, UK; $^{15}$  Departments of Physics and Astronomy, Univ. of
Toronto, Canada; $^{16}$ 
Istituto Nazionale di Geofisica, Roma, Italy}

\maketitle

\begin{abstract}
We report the first results from the BOOMERanG experiment, which
mapped at 90, 150, 240 and 410 GHz a wide ($3\%$) region of the 
microwave sky with minimal local contamination. From the data 
of the best 150 GHz detector we find evidence for a well defined peak in 
the power spectrum of temperature fluctuations of the Cosmic
Microwave Background, localized at $\ell = 197 \pm 6$, with an
amplitude of $(68 \pm 8) \mu K_{CMB}$. The location, width and amplitude 
of the peak is suggestive of acoustic oscillations
in the primeval plasma. In the framework of inflationary
adiabatic cosmological models the measured spectrum allows a Bayesian
estimate of the curvature of the Universe and of other cosmological
parameters. With reasonable priors we find $\Omega = (1.07 \pm 0.06)$ 
and $n_s = (1.00 \pm 0.08)$ (68$\%$C.L.) in excellent agreement with 
the expectations from the simplest inflationary theories. We also discuss the
limits on the density of baryons, of cold dark matter and on the cosmological
constant.

\end{abstract}

\section*{Introduction}

Acoustic oscillations in the primeval plasma before recombination ($z \sim
1100$)
produce a delicate pattern of horizon and sub-horizon size structures in
the
Cosmic Microwave Background. In the framework of inflationary adiabatic 
perturbations the angular power spectrum of these structures
features a harmonic series of peaks, with the first peak at multipole $\ell \sim
200$.

Here we present 
a high confidence measurement of the first peak resulting from the
analysis of the multiband, high quality image of about 3$\%$ of the sky 
obtained in the long duration flight of BOOMERanG \cite{deB00,Lan00}. 
We show that the BOOMERanG image is a faithful representation
of the CMB at angular scales smaller than 10$^o$, and we shortly discuss 
the methods we used to make sure that instrumental and astrophysical
contaminations are negligible in our results. We present the measured
power spectrum, and we discuss the cosmological implications of this
measurement.

\section*{The instrument and the data set}

The BOOMERanG payload and the important subsystems have been 
described in \cite{Mau97,Mas98,Mas99,Pia00,Cri00}.
It is a scanning telescope measuring simultaneously eight
pixels in the sky. Four pixels feature multiband photometers
(150, 240 and 410 GHz), two pixels have single-mode, diffraction 
limited detectors at 150 GHz and two pixels have single-mode, diffraction 
limited detectors at 90 GHz. The NEP of these detectors is below
200 $\mu K_{CMB} \sqrt{s}$ at 90, 150, and 240 GHz, and the angular
resolution
ranges from 10 to 18 arcmin FWHM. 
\begin{figure}[b!] 
\centerline{\epsfig{file=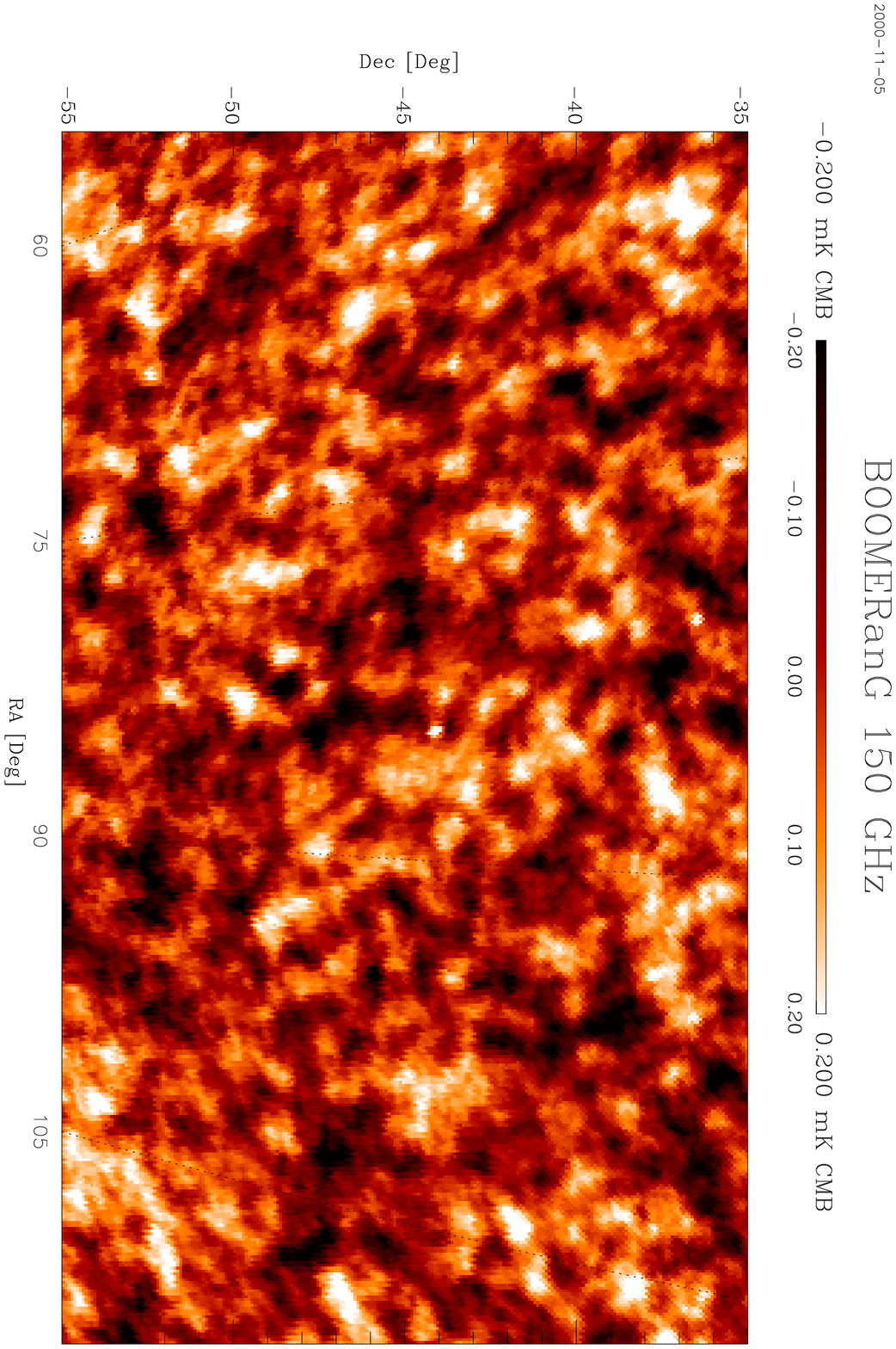,height=20cm,width=15cm}}
\vspace{10pt}
\caption{
The central part of the region observed by BOOMERanG at
150 GHz. The data have been high-pass filtered in the time domain to 
remove instrumental drifts. This process also removes structures in
the sky larger than 10$^o$. No further filtering is applied. The map
uses HEALPIX pixelization with 7' pixel side. Spectral and amplitude
arguments
show that the structures present in the map are degree-scale anisotropies
of the
Cosmic Microwave Background. 
}
\label{fig1}
\end{figure}
The continuous scan of the sky 
($\pm 30^o$ in azimuth at $1^o/s$ or $2^o/s$ ) 
allows a wide sky coverage. Moreover, it encodes different spherical
harmonic
components of the sky brightness into different frequencies within
a detection bandwidth ($\sim$ 0.2-20 Hz)
free from 1/f noise and drifts. The instrument is flown aboard a
stratospheric 
balloon  at 38 km of altitude, to avoid the bulk of atmospheric emission
and noise. 
Using a long duration balloon flight carried out by NASA-NSBF
around Antarctica we mapped $\sim 1800$ square degrees in the region
of the sky where Galactic contamination is minimal \cite{Sch98}.
The instrument collected 57 million samples for each detector
during the scans.  A pointing solution matrix (accurate to 3' rms) 
has been constructed using the data from the attitude sensors 
(sun sensors, laser gyroscopes, differential GPS). 
The beam response has been calibrated at ground, using a tethered
thermal source in the far field of the telescope. This calibration
has been checked in flight observing compact HII regions.
The responsivity has been calibrated against the
CMB dipole, which is visible along the scans as a $\sim$ 3 mK linear 
drift. Systematic effects limit the precision of this calibration to
10$\%$.
For further analysis the data have been high-passed in
the time domain, so that 1/f noise and drifts are reduced. 
In the same process, all the structures in the sky larger than
10$^o$ are effectively removed. A time-time correlation matrix has been
constructed
using an iterative method which effectively separates signal
and noise in the data \cite{Pru00}. The maximum likelihood map and its
covariance 
have been obtained from the data vector, the time-time correlation
matrix and the pointing matrix as described in \cite{Bor99}. We use
the HEALPIX \cite{Gor98} pixelization with 14' pixel side as a compromise 
between coverage of high multipoles and computation speed. 

\section*{The maps and the power spectra}

\begin{figure}[b!] 
\centerline{\epsfig{file=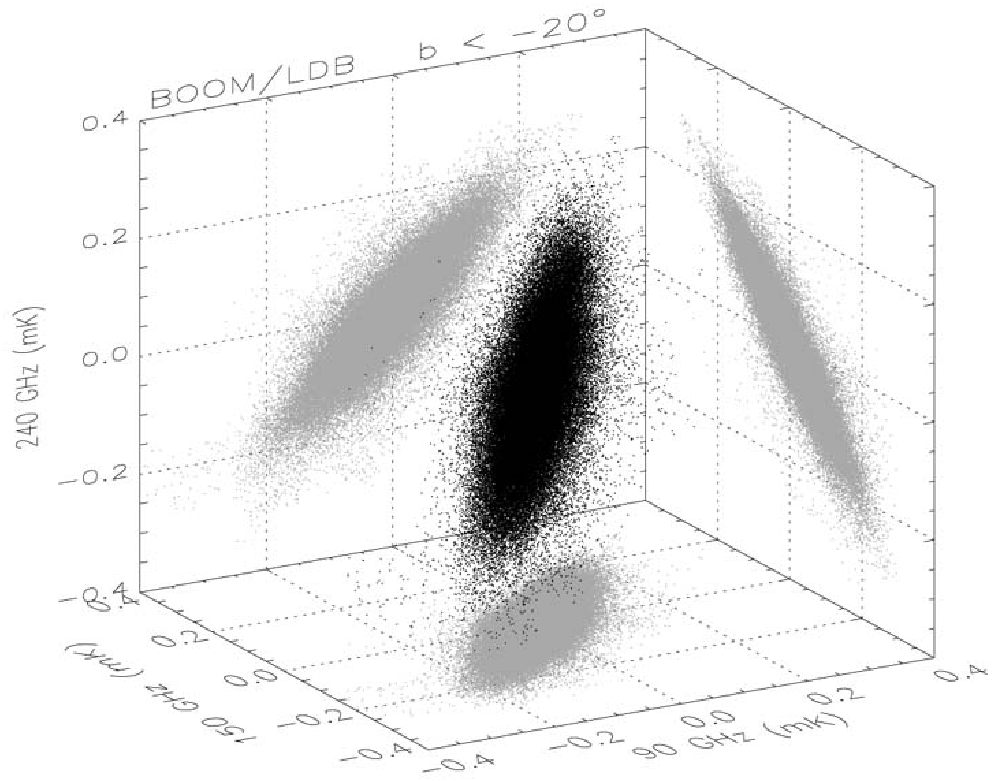,height=8cm,width=12cm}}
\vspace{10pt}
\caption{
3D and 2D scatter plots of the brightness measured by BOOMERanG at 
90, 150 and 240 GHz in each 7' pixel at high Galactic latitudes ($b <
-20^o$; 
84205 data). The units are mK$_{CMB}$, so that pure CMB fluctuations
produce 
2D scatter plots with slope 1.
}
\label{fig2}
\end{figure}

The maps in the four frequency channels have been published in
\cite{deB00}.
In figure \ref{fig1} we increase the contrast and show the central part of
a map obtained 
using a naive coadding in pixels of data from three of the 150 GHz
channels. The 90, 150, 240 GHz maps all present the same 
degree-size structures, with amplitude ratios characteristic of CMB
anisotropies
(see figure \ref{fig2}), thus demonstrating that the bulk of the
anisotropy
we measure is cosmological in origin. 
Dust contamination is evident only at lower Galactic latitudes, and mainly
in the
240 GHz channel. At high Galactic latitudes the 410 GHz channel is a 
good monitor of low level emission from interstellar
dust. We find a good correlation between this map and the IRAS/DIRBE
map extrapolated at 410 GHz (model 8 in ref. \cite{Sch98}) and filtered in
the time domain in the same way as our data. 
This correlation allows us to estimate low level contributions from 
IRAS-correlated dust to the CMB map shown in fig.1. We find (see
\cite{deB00,Mas00} for
details) that this dust can produce at most 2$\%$ of the mean square
anisotropy
plotted in the 150 GHz map. We also studied the effect of point sources
in the observed region. We used the PMN survey and the WOMBAT \cite{Wom00}
software
to extrapolate the radio sources flux to 150 GHz. The $rms$ brightness
fluctuation due to all
the sources in our field corresponds to $\sim 7 \mu K_{CMB}$, and has to
be compared
to the total measured sky fluctuation $\sim 80 \mu K_{CMB}$. Its
contribution to the
angular power spectrum has been computed and is reported in fig.3.
The maximum likelihood power spectrum of the central
part of the map (1$\%$ of the sky) is computed as described in
\cite{Bor99,deB00}
and is plotted in fig.3. A distinct peak at multipole $\ell \sim 200$
is
evident, and significant power is detected at all the multipoles up to the
highest observed ($\ell = 625$). 
\begin{figure}[b!] 
\centerline{\epsfig{file=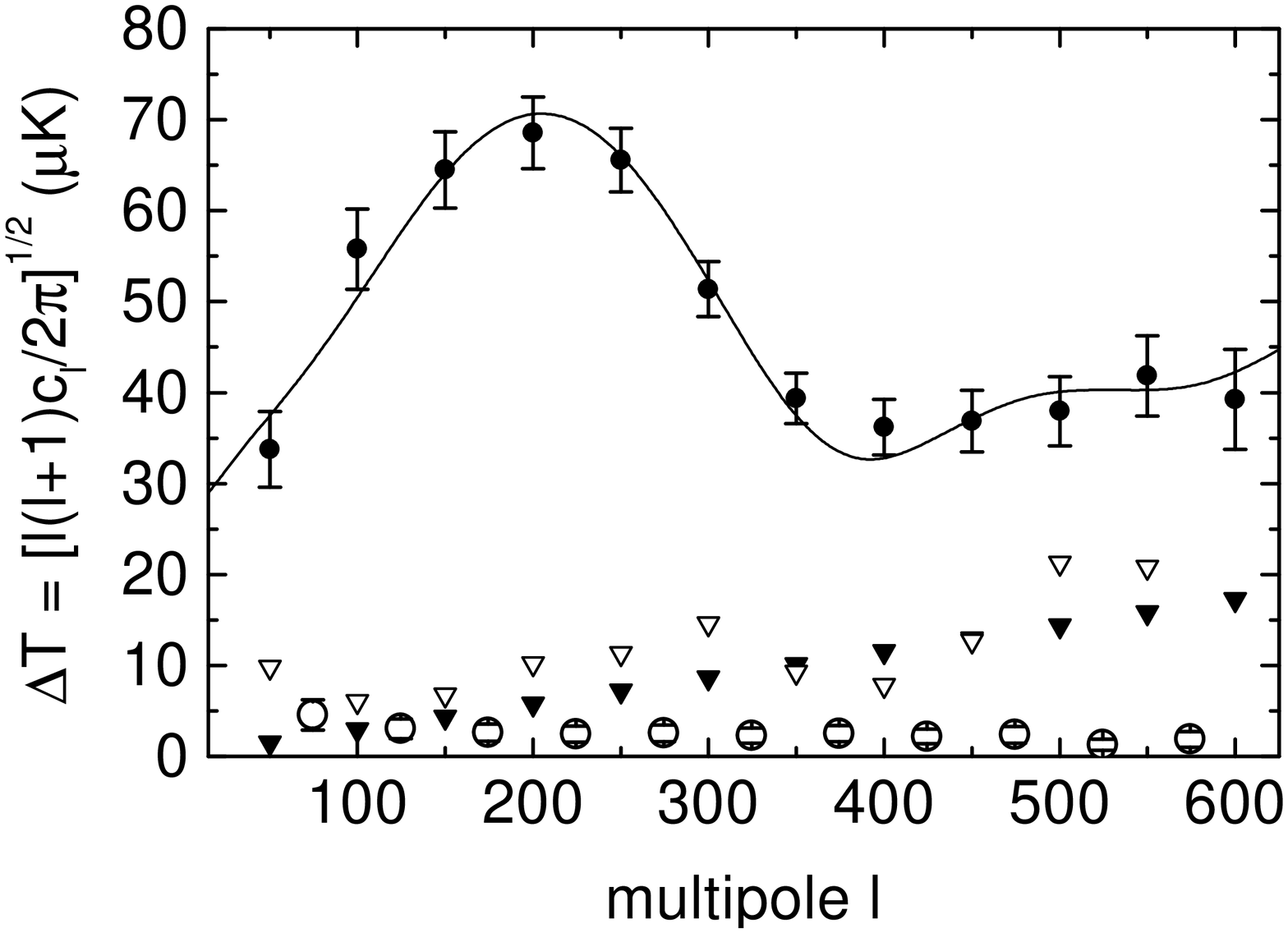,height=8cm,width=12cm}}
\vspace{10pt}
\caption{Power spectrum of 1$\%$ of the sky from one 150 GHz 
channel of BOOMERanG (filled circles). The continuous line is the
best fit adiabatic inflationary model. The open triangles represent
upper limits to systematic effects in the data; the filled
triangles are our estimate of the power from point sources in the
observed region; the open circles represent our estimate of the
power from IRAS-correlated galactic dust emission. Note that all these
fluctuations must be added in quadrature.}
\label{fig3}
\end{figure}
In figure \ref{fig4} we plot the {\it rms} signal in the center 
part of the maps at 90, 150, 240 and 410 GHz, computed as 
$\Delta T_{rms} = \sqrt{\sum_\ell (2 \ell + 1) c_\ell / 4 \pi}$.
The $\Delta T_{rms}$ have been converted into brightness
fluctuations using the measured spectral efficiency of the
different freqeuency channels. The $rms$ of the 410 GHz brightness
has been estimated from the component correlated with the 
IRAS/DIRBE maps; the amplitude of the thermal dust spectrum 
has been adjusted to fit such $rms$. The amplitudes of the other 
theoretical spectra for Galactic emission have been scaled
from \cite{DeO00} taking into account the different window functions
of BOOMERanG and COBE and assuming an $\ell^{-3}$ power spectrum. 
Once again it is evident that 
the data from the three channels more sensitive to the CMB are 
fit very well by the derivative of a 2.726K blackbody, while all the 
other reasonable spectra for millimeter wave emission originated 
in our galaxy cannot fit our measurements.
\begin{figure}[b!] 
\centerline{\epsfig{file=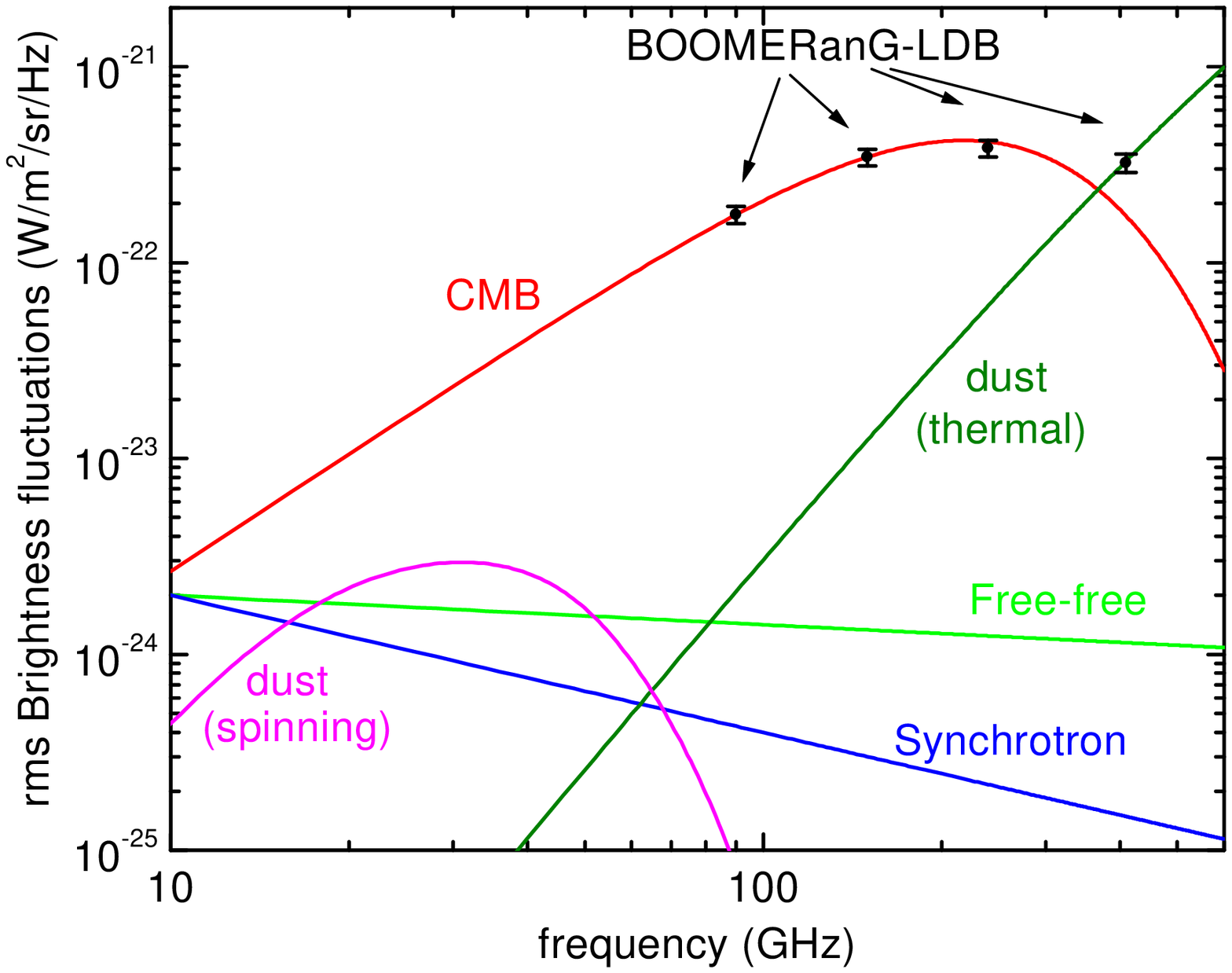,height=8.5cm,width=12cm}}
\vspace{10pt}
\caption{
rms brightness fluctuations measured by BOOMERanG
and spectra for CMB and local sources. The CMB curve has been 
computed from the derivative of a 2.726K blackbody. The thermal dust
curve has been normalized to the IRAS-correlated component detected
at 410 GHz in the BOOMERanG data. 
}
\label{fig4}
\end{figure}

\section*{Cosmological interpretation of the data}

Having been able to exclude all the suspect sources of local contamination
in the data of the power spectrum shown in figure 3, we 
can reasonably assume in the following that the detected power is
in the CMB and comes from structures at the recombination epoch ($z \sim
1100$). 
In the framework of adiabatic inflationary models, acoustic oscillations 
in the primeval plasma at horizon and sub-horizon scales produce 
a distinctive series of peaks in the power spectrum of the CMB. The
BOOMERanG data support this picture. The position, amplitude and 
width of the peak evident in fig.3 are consistent with the general
adiabatic inflationary scenario \cite{Hu97}, while the simplest models
based on
topological defects do not fit the data as well \cite{xxx}. 
Using a quadratic fit we find that the peak is located at multipole
$\ell = (197 \pm 6)$ (1$\sigma$). This result can shift a little bit if
one assumes skewed power spectra similar to the adiabatic scenario ones
\cite{Kno00}. This position is consistent with a flat geometry of space,
but 
it is not univocally related to the density parameter $\Omega$ if
we allow for a non vanishing cosmological constant \cite{Wei00}. The
continuous
line in fig.3, a very good fit to the measured data, has parameters
($\Omega$, $\Omega_b$, $\Omega_\Lambda$, $h$, $n_s$) = 
(1.035, 0.06, 0.425, 0.75, 1.0) where 
$\Omega_b$, $\Omega_\Lambda$, $\Omega$ 
are the density parameters for baryons, 
cosmological constant, and for the total mass-energy density 
respectively; $h$ is the dimensionless Hubble constant,
$n_s$ is the spectral index of the spectrum
of primordial density perturbations. A full Bayesian analysis has
been carried out \cite{Lan00} in order to measure the cosmological
parameters
and compute significant confidence intervals. In order to do this, 
we need to specify the prior distribution assumed for each of the
parameters. This is especially important in the case of CMB power spectrum
measurements, since an important geometrical degeneracy is present
\cite{Bon99}, 
so different combinations of the parameters produce very similar
power spectra. Given a flat model with $\Omega_\Lambda \sim 0.7$, 
a closed model producing the same power spectrum can be found 
decreasing $\Omega_\Lambda$ and $h$ and increasing $\Omega_b$, and 
an open model can be found doing the reverse. The
marginalized likelihood curve for $\Omega$ will then 
depend on the density of models
found in each direction along that particular path. 
Our 95$\%$ confidence intervals for $\Omega$ range from
$(0.88 - 1.12)$ to $(0.97 - 1.35)$ depending on the assumed priors
and parametrizations \cite{deB00,Lan00}. This strongly suggests a 
flat geometry of the Universe, and at least implies, with $95\%$
confidence,
a curvature length larger than 2 to 3 times the Hubble length today.
This measurement has very important cosmological consequences.
Several independent measurements point to a matter density parameter
significantly below unity \cite{Don99,Bah00,Bla99,Jus00,Wit00}.
As a consequence a different form of energy is needed to fill the gap 
between the total energy density of the Universe and the energy
density in matter. A cosmological constant is a good candidate,
and has the appealing feature of explaining the recent evidence 
coming from the observation of distant supernovae \cite{Rie98,Per99}
for an acceleration in the expansion rate of the Universe . The fine
tuning problem characteristic of the cosmological constant \cite{Wei89} 
can be overcome invoking quintessence
\cite{Ost95,Cal98,Arm00,Ame00,Bal200}.
In the same adiabatic perturbations framework, we constrain $n_s$
in the range $(0.71 - 1.01)$ to $(0.86 - 1.24)$ (95$\%$ confidence)
again depending on the priors \cite{Lan00}. Our constraints on $\Omega$
and $n_s$
are consistent and support the simplest inflationary scenarios
\cite{Lin81,Alb82,Wat00}. The third parameter constrained by BOOMERanG is
the
density of baryons. In the power spectrum 
this parameter controls the relative amplitude of the
first peak to the second one (see e.g. \cite{Hu97}). 
From the power spectrum of fig.4, this ratio
is larger than expected for a standard primordial nucleosynthesis 
($\Omega_b h^2 = (0.019 \pm 0.002)$, 1$\sigma$ \cite{Tyt00}), thus
suggesting
a physical density of baryons somewhat higher. Depending on the assumed
priors, we obtain
95$\%$ intervals for $\Omega_b h^2$ ranging from $(0.019 - 0.045)$ to 
$(0.026 - 0.048)$ \cite{Lan00}. This should be considered as a spectacular
agreement, since we are comparing the density of baryons inferred from the
physics of acoustic oscillations $\sim$ 300000 years after the big bang to
the density
of baryons inferred from the physics of nuclear reactions a few minutes
after
the big bang. Moreover, the measurement techniques are completely
orthogonal
and subject to completely different systematic effects. Should more
precise
measurements imply a more significant disagreement, new physics will be
needed,
and several hypothesis have been proposed already
\cite{aa,bb,cc,dd,ee,ff,gg}.
One way to improve the $\Omega_b$ consistency is to force the $n_s$ value
to be smaller. $n_s \sim 0.9$ is not the best marginalized value for 
$n_s$, but the fit to the data is very good and still consistent with
inflation 
(see e.g. \cite{Kin00}). In fig.5 we plot the BOOMERanG constraints in the
($\Omega_m$,$\Omega_\Lambda$) plane. Here $\Omega_m = \Omega_b +
\Omega_c$, 
with $\Omega_c$ density parameter for non relativistic dark matter.
It is evident that the data are fully 
consistent with a flat geometry of the Universe, represented by the 
$\Omega_m + \Omega_\Lambda = 1$ line; however, the CMB anisotropy data
alone 
cannot distinguish between the different contributions 
to the total mass-energy density. This degeneracy is efficiently removed
once
additional data are considered, like the data from distant supernovae or
the data
from the large scale distribution of galaxies. The combination of our data
with any of the two is quite powerful, and suggests a 'concordance' 
model with $\Omega_\Lambda \sim 0.7$ and $\Omega_m \sim 0.3$ (see \cite{Lan00} for
details). In ref. \cite{Teg00} all the available CMB power spectrum data 
and the power spectrum
of the distribution of IRAS galaxies from the redshift survey data are fit
simultaneously, confirming the results above.
\begin{figure}[b!] 
\centerline{\epsfig{file=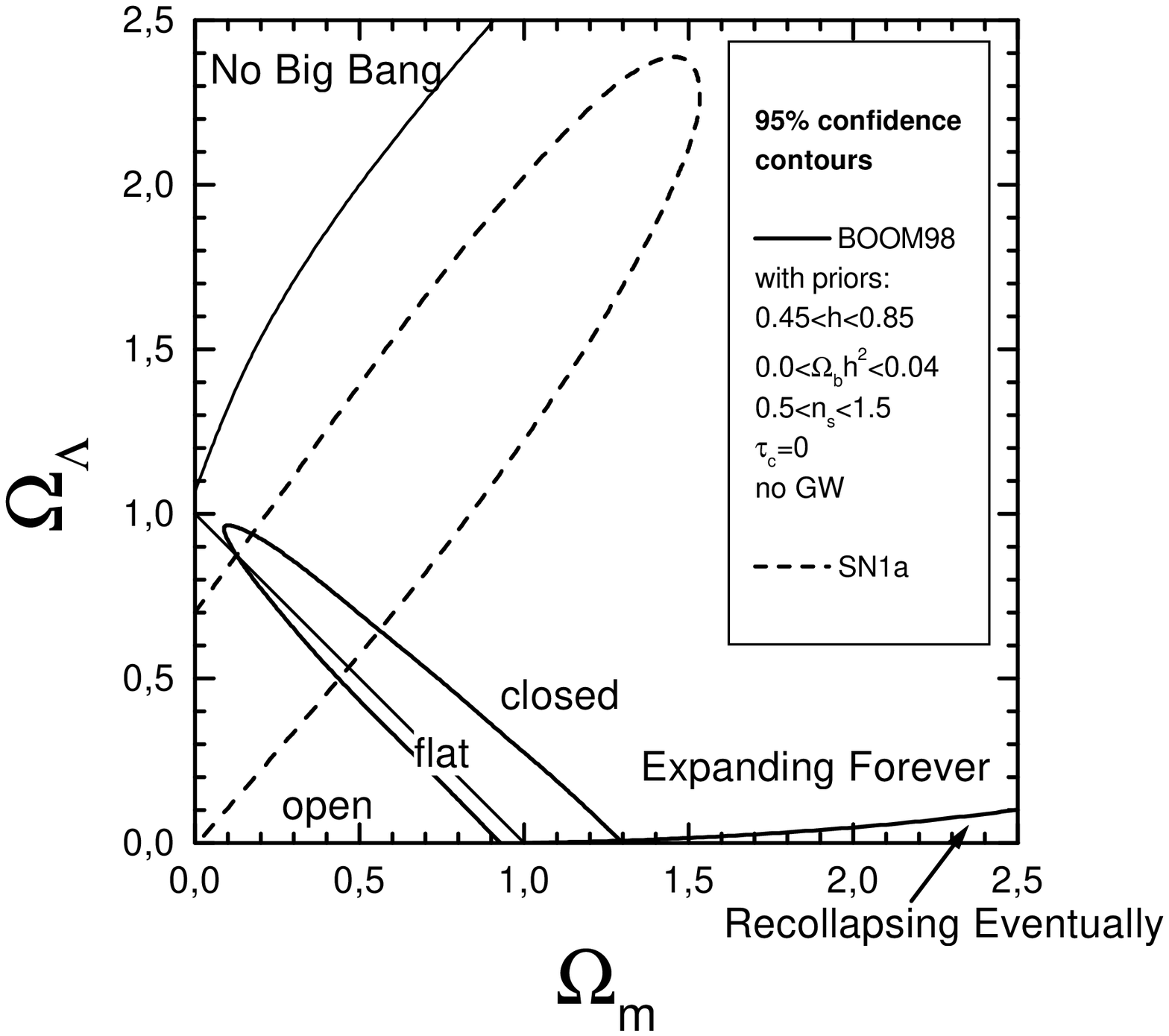,height=8.5cm,width=13cm}}
\vspace{10pt}
\caption{
95$\%$ confidence level constraints on cosmological parameters 
in the ($\Omega_m, \Omega_\Lambda$) plane, from the BOOMERanG power
spectrum (continuous contour) and from the high redshift supernovae
(dashed contour). 
}
\label{fig5}
\end{figure}

\section*{Conclusions}

BOOMERanG has produced a faithful, wide map of CMB anisotropy at angular
scales smaller
than 10$^o$. The power spectrum derived from the central part of the map
($\sim 1\%$ of
the sky) is consistent with the inflationary cosmological model with
gaussian adiabatic density fluctuations. A careful analysis of the power
spectrum
data, rigorously taking into account the priors and the degeneracies
related to these
parameters, has been carried out. We conclude that the geometry of our
universe is
close to flat, and that the primordial density flucutations have a nearly
Harrison-Zeldovich
power spectrum. The density of baryons inferred from these measurements 
is slightly higher than the standard 
nucleosynthesis value. Combining the BOOMERanG data with data from the
observations of
distant supernovae or from large scale structure studies we obtain
significant 
determinations of both $\Omega_m = (0.48 \pm 0.13)$ and $\Omega_\Lambda =
(0.66 \pm 0.07)$ (1$\sigma$). 
All these results come from a single detector, and use the central part of
the 
observed region. We are currently working on the combined analysis of all
the 
12 detectors sensitive to CMB. 

\section*{Acknowledgments}

The BOOMERanG project has been supported by PNRA, Universit\'a ``La
Sapienza'', 
and ASI in Italy, by NSF and NASA in the USA, and by PPARC in the 
UK. We would like to thank the entire staff of the NSBF, and the US 
Antarctic Program personnel in McMurdo for their excellent 
preflight support and a marvelous LDB flight. DoE/NERSC 
provided the supercomputing facilities. 
Web sites: (http://oberon.roma1.infn.it/boomerang) and 
(http://www.physics.ucsb.edu/$\sim$boomerang).

\end{document}